\documentclass[aps,floatfix,twocolumn,nofootinbib]{revtex4}

\usepackage{graphicx}
\usepackage{epic}
\usepackage{eepic}
\usepackage{latexsym}

\newcommand{\eq}[1]{(\ref{#1})}
\newcommand{\be}{\begin{equation}}
\newcommand{\ee}{\end{equation}}
\newcommand{\bea}{\begin{eqnarray}}
\newcommand{\eea}{\end{eqnarray}}

\newcommand{\hs}[1]{\hspace{#1 mm}}

\newcommand{\lf}{\left(}
\newcommand{\rg}{\right)}
\newcommand{\bk}{{\bf k}}
\newcommand{\bx}{{\bf x}}
\newcommand{\bkp}{{\bf k'}}
\newcommand{\bxp}{{\bf x'}}

\def\a{\alpha}
\def\b{\beta}
\def\cc{\gamma}
\def\C{\Gamma}
\def\d{\delta}
\def\D{\Delta}

\def\fr{\frac}

\def\m{\mu}
\def\n{\nu}

\def\r{\rho}
\def\s{\sigma}

\def\t{\tau}

\def\o{\omega}
\def\dg{\dagger}

\def\del{\partial}

\let\bm=\bibitem
\def\nn{\nonumber}

\begin{document}

\title{Fluctuations of Quantum Fields in a Classical Background and Reheating}

\author{Ali Kaya}
\email[]{ali.kaya@boun.edu.tr}
\affiliation{Bo\~{g}azi\c{c}i University, Department of Physics, \\ 34342,
Bebek, \.Istanbul, Turkey}

\date{\today}

\begin{abstract}
We consider the particle creation process associated with a quantum field $\chi$  in a time-dependent, homogeneous and isotropic, classical  background. It is shown that the field square $\chi^2$, the energy density and the pressure of the created particles have large fluctuations comparable to their vacuum expectation  values. Possible effects of these fluctuations on the reheating process after inflation are discussed. After determining  the correlation length of the fluctuations in two different models, corresponding to the decay  in the parametric resonance regime and in the perturbation theory, it is found that  these fluctuations should be taken into account in the final thermalization process, in  the back-reaction effects and when the formation of primordial black holes is considered.  In both models,  by comparing quantum and thermal fluctuations with each other it is observed that  very quick thermalization after the complete inflaton decay is not always possible even when the interaction rates are large. On the other hand, when the back-reaction effects are included during the preheating stage, the coherence of the inflaton oscillations is shown to be lost because of the  fluctuations in $\chi^2$. Finally, we note that a large fluctuation in the energy density may cause  a black hole to form and we determine the fraction of total energy density that goes into such primordial black holes in the model of preheating we consider.
\end{abstract}

\maketitle

\section{Introduction}

The theory of quantum fields in curved spacetime is by now a well established subject (see e.g. \cite{book1,book2,book3}). The issues like the non-uniqueness of the vacuum and the particle creation process, the renormalization of the stress-energy-momentum tensor are all thoroughly  understood, and the current research is mainly focused on the incorporation of  interactions and alternative rigorous definitions of the theory  (see e.g. \cite{cur1,cur2,cur3}). There are also some unsolved problems, the primary example being the black hole information paradox, but it seems that these can be explained in the full quantum theory of gravity and not in the semi-classical approximation which treats the background classically.  

Recently,  we have pointed out in \cite{m1,m2} an alternative description of the cosmological particle creation process, which tries to overcome two potential issues which may arise in the standard treatments based on the Fourier decomposition.  The first one is a (possible) conflict between causality, which requires the particle creation process to  take place independently in each Hubble volume, and the introduction of the globally defined  Fourier modes as the main physical observable. The second issue is related to a   subtlety in the interpretation of the vacuum expectation value of the number
operator of a momentum mode, since one can show that the number of created particles with a fixed momentum largely fluctuates about its mean value. To resolve these two issues, in \cite{m1,m2} we have analyzed the particle creation process using wave packets localized in a Hubble volume. As a result, the issue of causality is naturally settled by treating the creation of the modes in different Hubble volumes independently   and the mean values are interpreted as statistical averages over distinct  Hubble volumes. By applying  this formalism to reheating process after inflation, in  \cite{m1,m2} it is shown that there exists small density perturbations on Hubble length scales at the end of reheating. Although the
same result can also be obtained in the standard formulation  by introducing a suitable window function to probe the horizon scale, the construction presented in  \cite{m1,m2} is physically more direct and transparent.   

Since field theory respects locality by construction, the first issue mentioned above should not be a problem as long as the modes are physically interpreted properly. Indeed, the fact that field variables   must commute at {\it space-like} separations, i.e.  $[\chi(t',x'),\chi(t,x)]=0$ when $g_{\m\n}\D x^\m\D x^\n>0$, guarantees causality. The Fourier mode expansion of the field variable $\chi(t,x)$ is just a convenient way of describing physics. Any other complete set can be used for expansion and physics should not depend on the basis chosen. However, the second matter related to the interpretation of expectation values can be tricky, especially if there are large fluctuations about the mean values.  

In this paper, we consider the particle creation effects in a time-dependent, homogeneous and
isotropic, classical background and try to determine the magnitude of the
fluctuations of the  energy-momentum tensor $T_{\m\n}$ and the
field strength $\chi^2$ about their mean values. Namely, by defining  the
fluctuation operators   $\delta T_{\mu\nu}=T_{\mu\nu}-<T_{\mu\nu}>$ and
$\d_{\chi^2}=\chi^2-<\chi^2>$, whose expectation values vanish $<\d
T_{\mu\nu}>=<\d_{\chi^2}>=0$, we calculate the variances $<(\d T_{\m\n})^2>\not
=0$ and $<(\d_{\chi^2})^2>\not =0$. As we will see in section \ref{ch2}, for
most of the  components of the energy-momentum tensor including the energy
density and the pressure, and for $\chi^2$ the deviations have the same order of
magnitude as the corresponding mean values, which shows the existence of large
fluctuations. We emphasize that the spatial scale of these fluctuations is given
by the correlation length of the quantum field $\chi$. 

In the second part of this paper, in section  \ref{ch3}, we discuss possible
implications of these results for the reheating process in single scalar field
inflationary models. We consider the decay of the
inflaton field $\phi$ into bosonic $\chi$ particles in two different prototype 
models corresponding to perturbative and  parametric resonance regimes,
and in both models we determine the correlation length of the $\chi$
field excited by the inflaton oscillations. We find that the existence
of these fluctuations may affect various events during reheating. It
is pointed out  in section \ref{ch3} that thermalization process cannot be
completed unless quantum fluctuations become smaller than the thermal
fluctuations in equilibrium and this may delay the final moment of  reaching
thermal equilibrium. Secondly, in the model with preheating, when the
$\chi$ field starts affecting the frequency of the background 
inflaton oscillations, the fluctuations in $\chi^2$  are found to destroy the
coherence of the oscillations. We show that because of this loss of coherence it is
no longer possible to neglect the spatial derivatives of the inflaton when
the back-reaction effects become important. Finally in section
\ref{ch3}, the possibility of black hole formation  because of  the fluctuations in the
energy density of created particles is considered. For a large fluctuation,
which of course happens rarely, the corresponding Schwarzschild radius may become
larger than the correlation length which would produce a black hole.  We show
that the observational constraint on the fractional energy density that can go
into primordial black holes puts some new restrictions, 
especially in models of preheating. We finally conclude by reviewing our
findings and discussing future directions in section \ref{ch4}. 
 
\section{The fluctuations} \label{ch2}

We consider a real  scalar field $\chi$ propagating in a cosmological Robertson-Walker  metric  
\be\label{m}
ds^2=-dt^2+a^2(dx^2+dy^2+dz^2),
\ee
which has the following action
\be\label{ac1}
S=-\fr12 \int \sqrt{-g}\left[(\nabla\chi)^2+M^2 \chi^2\right].
\ee
We assume that in addition to the scale factor $a$, the mass parameter $M$ may
also depend on time: $M=M(t)$. Therefore,  particle creation can be induced both
by the evolution of the metric and by the externally varying time dependent
mass parameter $M$. The energy-momentum tensor can be obtained from \eq{ac1} as 
\be\label{em}
T_{\mu\nu}=\nabla_\mu\chi\nabla_\nu\chi-\fr{1}{2}g_{\mu\nu}\left[
\nabla_\b\chi\nabla^\b\chi+M^2\chi^2\right].  
\ee
More explicitly, one can read the energy density $\r$, the pressure $P$, the "momentum" $U_i$ and (trace-free) stress components $\tau_{ij}$ as 
\bea
\r&=&\fr{1}{2}\dot{\chi}^2+\fr{1}{2}g^{ij}(\del_i\chi)(\del_j\chi)+\fr{1}{2}
M^2\chi^2,\nn\\
P&=&\fr{1}{2}\dot{\chi}^2-\fr{1}{6}g^{ij}(\del_i\chi)(\del_j\chi)-\fr{1}{2}
M^2\chi^2,\nn\\
U_i&\equiv&T_{ti}=\dot{\chi}(\del_i\chi)=\fr{1}{2}\left[\dot{\chi}
(\del_i\chi)+(\del_i\chi)\dot{\chi}\right],\label{ex}\\ 
\tau_{ij}&=&\del_i\chi\del_j\chi-\fr{1}{3}g_{ij}\left[g^{kl}
(\del_k\chi)(\del_l\chi)\right],\nn
\eea
where dot denotes time derivative and all indices refer to the obvious
coordinate basis of \eq{m}.
Note that the spatial components of the energy-momentum tensor are decomposed as
$T_{ij}=\tau_{ij}+Pg_{ij}$, where $g^{ij}\t_{ij}=0$. In quantum theory, there is
an ordering ambiguity in
$U_i$ and in \eq{ex} we perform a symmetric ordering.  

For quantization it is convenient to define a new field $X$ by  
\be\label{y}
X=a^{3/2} \chi.
\ee
Then, the action \eq{ac1} up to surface terms becomes 
\bea\nn
S=\fr12 \int \left[\dot{X}^2-g^{ij}(\del_i X)(\del_jX)-(M^2-\fr94 H^2 -\fr32
\dot{H}) X^2\right],
\eea
where $H=\dot{a}/a$ is the Hubble parameter. The momentum variable $\Pi$
conjugate to $X$ is given by $\Pi=\dot{X}$ and one can easily apply the standard
canonical quantization procedure. Introducing the mode function $X_k$  and {\it time
independent}  ladder operator corresponding to a fixed time $t_0$, $a_\bk$, as  
\bea\label{exp}
X=\fr{1}{(2\pi)^{3/2}} \int d^3 k\left[ a_\bk X_k e^{-i\bk.\bx}+  a_\bk^\dagger
X_k^* e^{i\bk.\bx}\right], 
\eea
the canonical commutation relation $[X(t,\bx),\Pi(t,\bxp)]=i\d(\bx-\bxp)$  can be satisfied by imposing  
\be\label{fcc}
[a_\bk,a^\dagger_\bkp]=\d(\bk-\bkp),
\ee
and 
\be
X_k\dot{X}_k^*-X_k^*\dot{X}_k=i,
\ee
where boldface letters $\bk$ and $\bx$ refer to the spatial 3-vectors $k_i$ and
$x^i$, $\bk.\bx=k_ix^i$ and $k^2=\d^{ij}k_ik_j$. On the other hand, the field equations imply
\be\label{xeq}
\ddot{X}_k+\o_k^2X_k=0, 
\ee
where 
\be\label{o}
\o_k^2=M^2+\fr{k^2}{a^2}-\fr94 H^2 -\fr32 \dot{H}.
\ee
Because of the  homogeniety and isotropy of the background, the mode function
$X_k$ depends only on the magnitude $k$ of $\bk$.   

The ground state of the system at time $t_0$ can be defined as
\be\label{gs}
a_\bk|0>=0,
\ee
and in terms of the mode functions this corresponds to the initial conditions 
\be\label{12}
X_k(t_0)=\fr{1}{\sqrt{2\o_k}},\hs{3}\dot{X}_k(t_0)=-i\sqrt{\fr{\o_k}{2}}.
\ee
Using \eq{12}, the Hamiltonian at time $t_0$ can be found as 
\be
H(t_0)=\int d^3 k\left[a_\bk^\dg a_\bk +\fr{1}{2}\right]\,\o_k,
\ee
which justifies the identification of  $|0>$ defined in \eq{gs} as the ground
state. Let us note that the above formulation is identical to the standard
formulation of particle creation in terms of Bogoliubov transformations (see \cite{parker}). 
The correspondence can be achieved by defining the time dependent  $\a_k$ and
$\b_k$ coefficients as 
\be\label{ab}
X_k=\fr{1}{\sqrt{2\o_k}}\left[\a_k\,e^{-i\int_{t_0}^t\o_k
dt'}+\b_k\,e^{i\int_{t_0}^t\o_k dt'}\right].
\ee
By imposing the following  equations 
\bea
\dot{\a}_k&=&\fr{\dot{\o_k}}{2\o_k}e^{2i\int^t_{t_0} \o_k dt}\b_k,\nn\\
\dot{\b}_k&=&\fr{\dot{\o_k}}{2\o_k}e^{-2i\int^t_{t_0} \o_k dt}\a_k, \nn
\eea
and the initial conditions $\a_k(t_0)=1$, $\b_k(t_0)=0$, the equivalence of both
formulations can easily be established. For future use let us also calculate $\dot{X}_k$ as 
\be\label{ab2}
\dot{X}_k= \sqrt{\fr{\o_k}{2}}\left[-i\a_k\,e^{-i\int_{t_0}^t\o_k
dt'}+i\b_k\,e^{i\int_{t_0}^t\o_k dt'}\right]. 
\ee
Note that because of the equations obeyed by  $\a_k$ and $\b_k$, in taking time
derivative of $X_k$ from \eq{ab} one can treat $\a_k$ and $\b_k$  as if they are constant
parameters. 

The particle creation process is usually described by giving  the vacuum
expectation value of the number operator for a mode of momentum $k$.
Alternatively, one can also calculate the vacuum expectation value of the energy-momentum tensor $<T_{\m\n}>\equiv<0|T_{\m\n}|0>$. One benefit of working with the energy-momentum tensor is that  the potential problem mentioned in the introduction,
namely the creation of a global Fourier mode in an
expanding universe, is naturally evaded. Using \eq{exp} in \eq{ex}, one can
straightforwardly calculate 
\bea
<\r>&=&T+V+G,\nn\\
<P>&=&T-V-\fr{1}{3}G\label{vev},\\
<U_i>&=&0,\hs{3}<\t_{ij}>=0,\nn
\eea
where being the real functions of time $T$, $V$ and $G$ are given by
\bea
T&=&\fr{1}{2(2\pi a)^3}\int d^3 k\, |\dot{X}_k-\fr{3}{2}HX_k|^2,\nn\\
V&=&\fr{1}{2(2\pi a)^3}\int d^3 k\, M^2\, |X_k|^2,\label{func}\\
G&=&\fr{1}{2(2\pi a)^3}\int d^3 k \,\fr{k^2}{a^2}\,|X_k|^2.\nn
\eea
Here,  $T$ and $V$ play the roles of the kinetic and potential energies,
respectively, and $G$ is like the energy stored in the gradient of the field. 
It is an easy exercise to show that when $M$ is constant 
the expectation value of the energy-momentum tensor is conserved 
\be\label{cons}
M(t)=M_0\,\Rightarrow\,  \nabla_\m<T^{\m\n}>=0,
\ee
which actually  reduces to a single nontrivial equation
$\dot{<\r>}+3H(<\r>+<P>)=0$.  Thus, in that case $<T_{\m\n}>$  can be fed into the right hand
side of the Einstein's equations to include the back-reaction effects. 

The functions $T$, $V$ and $G$ defined in \eq{func} diverge in general.
Therefore, they should be regularized to be interpreted physically. Adiabatic
(or WKB) 
regularization is a suitable and physically appealing way of getting finite
expectation
values for  $<\chi^2>$, $<\r>$ and $<P>$ \cite{ad1,ad2}. One can see
that the function $V$ is related to $<\chi^2>$ and thus it can be made finite by
adiabatic regularization. Similarly, the functions $T$ and $G$ can also be
expressed in terms of $<\r>$, $<P>$ and 
$<\chi^2>$, thus they can also be regularized.  When we mention about  the
magnitudes of the variables $T$, $V$ and $G$ below, we implicitly refer to their
regularized finite values.  

The fact that $<U_i>=0$ and $<\t_{ij}>=0$ does not imply $U_i=0$ and $\t_{ij}=0$
identically. Similarly, $\r$ and $P$ do not exactly equal their mean values.
All these physical quantities fluctuate about their expectation values and we
would like to find out the size of  these fluctuations. For that we define the
fluctuation operator 
\be
\d T_{\m\n}=T_{\m\n}-<T_{\m\n}>,
\ee
and try to compute the variance $<(\d T_{\m\n})^2>\not=0$.  

Let us start with the energy density $\d\r=\r-<\r>$. Using \eq{exp} in the
definition of $\r$ given in \eq{ex} and reading $<\r>$ from \eq{vev}, a
relatively long but straightforward calculation gives 
\bea
<\d\r^2>&=&2T^2+2V^2+\fr{2}{3}G^2   \label{dr}\\
&+&\fr{M^2}{(2\pi a)^6}\left[\int d^3 k
\cos(\phi_k)|\dot{X}_k-\fr{3}{2}HX_k||X_k|\right]^2\nn\\
&-&\fr{M^2}{(2\pi a)^6}\left[\int d^3 k
\sin(\phi_k)|\dot{X}_k-\fr{3}{2}HX_k||X_k|\right]^2,\nn
\eea
where $\phi_k$ is the phase difference between $(\dot{X}_k-\fr{3}{2}HX_k)$ and
$X_k$. The only "nontrivial" step in this calculation is to deal with a double
integral which has the following form  
\be
\int\int d^3 k_1\, d^3 k_2\, (\bk_1.\bk_2)^2\,  |X_{k_1}|^2 |X_{k_2}|^2.
\ee
This can be converted by suitable angular integrations into   
\be
\fr{1}{3}\left[\int d^3 k_1 (k_1^2)\,|X_{k_1}|^2\right] \left[\int  d^3 k_2\,
(k_2^2)|X_{k_2}|^2 \right], 
\ee
which can then be expressed in terms of the function $G$. As a check on this
computation one can see from \eq{func} and \eq{dr} that  
\bea
<\d\r^2>&\geq& 2T^2+2V^2\nn\\
&-&\fr{M^2}{(2\pi a)^6}\left[\int d^3 k
|\dot{X}_k-\fr{3}{2}HX_k||X_k|\right]^2\label{ineq}\\
&=&\fr{1}{2(2\pi a)^6}\int d^3k_1d^3k_2\left[  a_{k_1}
a_{k_2}-b_{k_1}b_{k_2}\right]^2\geq 0,\nn
\eea
where $a_k=|\dot{X}_k-\fr{3}{2}HX_k|$ and $b_k= M|X_k|$. This shows that
$<\d\r^2>\geq0$, as it should be.  

Let us now turn to the fluctuations of the pressure and define  $\d P=P-<P>$. Again a
straightforward calculation, which uses \eq{ex}, \eq{exp} and \eq{vev}, gives   
\bea
<\d P^2>&=&2T^2+2V^2+\fr{2}{27}G^2   \label{dp}\\
&-&\fr{M^2}{(2\pi a)^6}\left[\int d^3 k
\cos(\phi_k)|\dot{X}_k-\fr{3}{2}HX_k||X_k|\right]^2\nn\\
&+&\fr{M^2}{(2\pi a)^6}\left[\int d^3 k
\sin(\phi_k)|\dot{X}_k-\fr{3}{2}HX_k||X_k|\right]^2.\nn
\eea
Here also one has $<\d P^2>\geq0$, as expected.

Both in \eq{dr} and in \eq{dp}, there are integrals involving the phase
$\phi_k$, which cannot be expressed in terms of the functions $T$, $V$ and $G$.
Interestingly, however, these terms appear with opposite signs in $<\d\r^2>$ and
$<\d P^2>$, and from \eq{dr} and \eq{dp} one sees that  
\be\label{main1}
<\d\r^2>+<\d P^2>=4T^2+4V^2+\fr{20}{27}G^2.
\ee
Comparing with the average $<\r>$ given in \eq{vev},   eq. \eq{main1} shows that
$<\d\r^2>+<\d P^2>$ has the same order of magnitude as the mean energy density
square
\be
 <\d\r^2>+<\d P^2>=c (<\r>)^2,
\ee
where $c$ is a number close to unity, which varies depending on the hierarchy
between the
functions $T$, $V$ and $G$. For instance, $c\sim 32/27$ if $T\sim G\gg V$, or
$c\sim 0.97$ if  $T\sim V\sim G$.  

Before proceeding with the calculations of other fluctuations, let us make a few
comments. It is easy to see from \eq{ineq} that the integrals
involving $\cos(\phi_k)$ and $\sin(\phi_k)$ in \eq{dr} and \eq{dp} are bounded
by $2T^2+2V^2$. Therefore, one would expect them to be regularized in a suitable
scheme. On the other hand, the sum of the fluctuations $<\d\r^2>+<\d P^2>$
can already be expressed in terms of {\it finite} functions $T$, $V$ and $G$ after
adiabatic regularization. In general, one would also expect the oscillating
$\cos(\phi_k)$ and $\sin(\phi_k)$ integrals in \eq{dr} and \eq{dp} to be smaller than the
previous terms. Therefore, both $<\d\r^2>$ and $<\d P^2>$ should have the same order of
magnitude as $(<\r>)^2$. In any case, \eq{main1} shows that the
fluctuations $<\d\r^2>$ and $<\d P^2>$ cannot be  {\it simultaneously} small
compared to the average energy density. This can be viewed as  an uncertainty relation
between the energy density  and pressure fluctuations. 

Because of these large fluctuations, it may not always be possible
to use a simple equation of state $P=w \r$ to characterize the energy-momentum
tensor in quantum particle creation. One can calculate the correlation of the 
energy and pressure 
fluctuations  as 
\bea
<\{\d\r,\d P\}>=4T^2-4V^2-\fr{4}{9}G^2,\nn
\eea
where the symmetric ordering  $\{\d\r,\d P\}=\d\r\d P+\d P\d\r$ is chosen inside
the brackets. Thus, depending on the case the correlation can be negligible,
which would forbid the 
use of an effective equation of state parameter $w$. One can also determine the
expectation value of the commutator, which should give information about simultaneous
measurability of fluctuations. We find  
\bea
<[\d P,\d \r]>=\fr{4M^2}{(2\pi a)^6}\left[\int d^3 k
\cos(\phi_k)|\dot{X}_k-\fr{3}{2}HX_k||X_k|\right]\nn\\ 
\times\left[\int d^3 k \sin(\phi_k)|\dot{X}_k-\fr{3}{2}HX_k||X_k|\right],\nn
\eea
therefore the integrals involving the phase $\phi_k$ is related to the
expectation value of the commutator. Note that the expectation value is equal to
zero for $M=0$.  

Let us now continue with the other components of the energy-momentum tensor,
namely $U_i$ and $\t_{ij}$. As noted in \eq{vev}, their expectation values
vanish. To determine fluctuations, one can calculate  
\be
<U_iU_j>=\fr{4}{3}T\,G\,g_{ij}.
\ee
and
\be
<\t_{ij}\t_{ij}>=\begin{cases}{\fr{4}{9}G^2\hs{5} \textrm{ if}\hs{3} i\not=j,
\cr \cr
                                \fr{16}{27}G^2\hs{5} \textrm{if}\hs{3}
i=j,}\end{cases}
\ee
where in the last equation the summation convention is not used, i.e. $i$ and $j$ are treated as free indices.  The correlation of field variables carrying different set of indices can be seen to vanish because of the homogeneity and isotropy of the background. However, when $G$ has the same order of magnitude as $T$ and $V$, the fluctuations of the stress components $\t_{ij}$ cannot simply be ignored.  Remember  that $G$ is the energy stored in the gradient of the field, therefore it is not surprising to discover that the fluctuations of the stress components depend on $G$. Similarly, for large $G$ and $T$, the "momentum" components $U_i$ has large fluctuations, of the order of average energy density. Again it is natural to see both $T$ and $G$ in the  fluctuations of momentum, since $T$ measures kinetic energy and $G$ measures spatial variations.  

When back-reaction or symmetry breaking effects are considered, $\chi^2$ becomes an important parameter and thus it is crucial to determine  the fluctuations about the mean $<\chi^2>$. Defining as before the variation operator  $\d_{\chi^2}=\chi^2-<\chi^2>$ and using \eq{y} and \eq{exp}, one can easily calculate 
\be\label{g}
<(\d_{ \chi^2})^2>= 2 (<\chi^2>)^2. 
\ee
Therefore, viewing $\chi^2$ as a random variable one sees that  the
corresponding standard deviation is equal to $\sqrt{2}$ times the average,
which again indicates the existence of large fluctuations about the mean. This
result can indeed be anticipated without doing any computation since the free field
$\chi(t,\bx)$ can be viewed as a collection  Gaussian random variables defined  at each point in space  and \eq{g} is true for any Gaussian distribution.  

\begin{figure}
\centerline{
\includegraphics[width=7cm]{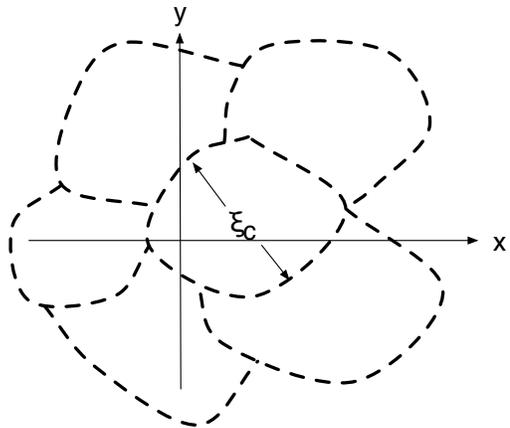}}
\caption{The space divided into regions of typical (comoving) size $\xi_c$. At a given instant, the $\chi$ field  can be assumed to vary appreciably from one region to another, while it is nearly uniform in each region.}
\label{fig1}
\end{figure}

Let us remind that the expectation values of physical quantities determined
above  should be interpreted as statistical averages over different points in
space at a given  time. However, the field variables at nearby points are not independent, i.e. they are
correlated with each other. The size of a region containing such correlated
field variables is given by the {\it correlation length} $\xi_c$, which can be determined from
the two point function
\bea
<\chi(t,\bx)\chi(t,0)>=\fr{1}{(2\pi a)^3}\int d^3 k\, |X_k|^2 e^{i\bk.\bx}\nn\\
=\fr{1}{2\pi^2 a^3}\int_0^\infty  dk\,k^2
|X_k|^2\left[\fr{\sin(kr)}{kr}\right],\label{cl}
\eea
where $r^2=\d_{ij}x^ix^j$. The comoving correlation length $\xi_c$ is the minimum value of $r$ that makes \eq{cl} to vanish. Physically, it gives the spatial comoving size of a region in which the $\chi$ field is homogenous.  Similarly, it is also possible to define the correlation lengths of the energy density, the pressure etc., however these are expected to have the same order of magnitude as $\xi_c$. Therefore,
one can imagine the space to be divided into comoving regions of typical size $\xi_c$, and in each
region the physical quantities like the $\chi$ field, the energy density $\r$ and the
pressure $P$ are uniformly distributed. The fluctuations, on the other hand, give variations from
one region to another (see figure \ref{fig1}).  

As $r\to 0$,  \eq{cl} should give the (regularized) mean value $<\chi^2>$. Therefore, in calculating the correlation length from \eq{cl}, a suitable regularization should be performed  before the momentum integral is taken (note that \eq{cl} is finite for $r\not =0$ without any need of regularization). To employ WKB regularization \cite{ad2} one uses \eq{ab} in \eq{cl} and sets $|\a_k|^2=|\b_k|^2+1$, which yields   
\bea
&&<\chi(t,\bx)\chi(t,0)>=\fr{1}{2\pi^2 a^3}\int_0^\infty  dk\,\fr{k^2}{\o_k}
\times\label{c}\\
&&\left[ |\b_k|^2+\textrm{Re}\lf\a_k\b_k^* e^{-2i\int_{t_0}^t \o_k dt'}\rg
\right]\left[\fr{\sin(kr)}{kr}\right],\nn 
\eea
where an additive  factor of  $1/2$ is ignored in the first square brackets for regularization.  Since initially one imposes $\b_k(t_0)=0$, the momentum integral in \eq{c} equals zero at $t=t_0$ and it is expected to converge also at later times. We will use this regularization in determining the correlation
length in the next section.  

Let us note that in our calculations we ignore the back-reaction effects and assume that the evolution of the background does not change as the  particles are created. This is an approximation and, for instance, as the energy of the created particles increases the energy of the background should decrease. Naively, one may then tend to view these  fluctuations as "isocurvature" perturbations, however, this depends on how the energy density of the background and the created particles redshift. On the other hand, keeping in mind that  the back-reaction effects are ignored may help to solve the following concern. As noted in \eq{cons}, when $M$ is constant  the average of the energy-momentum tensor is conserved and one may worry that this can be spoiled  if fluctuations are added on top of the mean values. However, all that is needed is to satisfy the energy-momentum conservation for the whole system, and this should be guaranteed when the back-reaction effects are properly taken into account. 

It is interesting to compare the direct evaluation of the dispersion used in this paper with the standard formulation which employs  a smearing of  the dispersion (or power spectrum) in momentum space with a suitable window function to probe a given scale. In general these two computations should agree, at least on large scales, however there are some important differences.  While using a window function can be a convenient way of  regularizing some expressions,  it is not possible to identify the correlation length, since the power spectrum does not contain any information about the correlation of variables at different points. Moreover, some results may be sensitive on the choice of a window function whereas the  adiabatic regularization offers both technically and physically unique way  of getting finite and consistent values from singular expressions.  To emphasize the difference, let us point out that recently in \cite{ad3,ad4} adiabatic regularization is applied to cosmological perturbation theory and some standard results, such as  the form of the power spectrum, are shown to be modified significantly depending on the regularization. 

\section{Implications for reheating}\label{ch3}

In this section, we analyze  the particle creation process in the reheating
period  after inflation and consider the single scalar field inflationary models. In general, 
the evolution of the metric and the inflaton $\phi$ is governed by the Friedmann  
and the scalar field equations  
\bea
&&H^2=\fr{8\pi}{3M_p^2}\left[\fr12 \dot{\phi}^2+V(\phi)\right],\nn\\
&&\ddot{\phi}+3H\dot{\phi}+\fr{\del V}{\del\phi}=0,\label{fe2}
\eea
where $V(\phi)$ is the scalar potential. In reheating, the scalar oscillates
about the minimum of the potential and we focus on models in which the potential
at that stage can be taken as  
\be\label{pot}
V=\fr12 m^2\phi^2,
\ee
where $m$ is the inflaton mass. Because of the expansion of the universe, the
amplitude of the oscillations slowly decreases in time so one can assume a
solution of the form 
\be
\phi=\Phi(t)\,\sin(mt).
\ee
When $\dot{\Phi}\ll m\Phi$, the field equations  \eq{fe2} become 
\be\label{sol}
H^2=\fr{4\pi m^2}{3M_p^2} \Phi^2,\hs{5}\dot{\Phi}+\fr32 H\Phi=0,
\ee
and these can be solved as
\be\label{sol1}
a=a_0\left(\fr{t}{t_0}\right)^{2/3},\hs{5}\Phi=\fr{M_p}{\sqrt{3\pi}mt}.
\ee
Thus, the evolution is equivalent to the dust dominated universe and
in that case the combination $9H^2/4+3\dot{H}/2$, which appears in \eq{o}, is
equal  to zero.  

We assume that reheating occurs due to the coupling of the inflaton to a bosonic field, $\chi$, and  consider two different types of interactions suitable for preheating and perturbative decay, respectively. We first determine the correlation length $\xi_c$ of the excited $\chi$ field at the end of decay and evaluate the functions  $T$, $V$ and $G$ introduced in the previous section in \eq{func}. Later we study possible implications of our findings on the  three important processes,  which are  the final thermalization of the decay products, the back-reaction effects and the formation of primordial black holes.  

\subsection*{Preheating}

Let us first start with preheating. The decay of the inflaton field in the
parametric resonance regime has been analyzed in detail in
\cite{pr1,kls1,pr2,kls2}. Specifically, it is shown in  \cite{kls1,kls2} that 
the
following interaction 
\be\label{int1}
{\cal L}_{int}=-\fr12 g^2 \phi^2\chi^2,
\ee
can give a decay due to broad parametric resonance if the parameter
$q=g^2\Phi^2/m^2\gg1$. This condition can naturally be satisfied
in chaotic inflationary scenario in which $\Phi_0\sim M_p$, where $\Phi_0$ is
the initial value of the inflaton amplitude and $M_p$ is the Planck mass. In
this model the
frequency \eq{o} becomes  
 \be\label{om1}
\o_k^2=\fr{k^2}{a^2}+g^2\Phi^2\sin^2(mt),
\ee 
and the particle creation mainly  occurs due to the oscillating term.  As shown
in \cite{kls1,kls2}, for momenta in certain instability bands, the solution of
\eq{xeq} exponentially grows, which corresponds to 
$|\a_k|\simeq|\b_k|\gg1$. Actually in the expanding universe the
real process is much more complicated where a given momentum, which is
initially in the first band,  jumps over many different instability bands in
time. Fortunately, it  can still be treated analytically by approximating the
process in terms of successive scattering on parabolic potentials \cite{kls2}.
The final result for the Bogoliubov coefficients can effectively be described by
an index $\m_k$ such that 
\be\label{b}
|\a_k|\simeq |\b_k|=e^{\m_k m t}.
\ee
Moreover, the first and the most important resonance band is initially peaked
around the {\it physical momentum}  $k_*=\sqrt{gm\Phi_0}$, with a width of the order
of $k_*$ (here we only consider the first stage of preheating where
back-reaction and re-scattering effects are ignored). The effective index
$\m_k$ can be approximated as 
\be\label{ma}
\m_k\simeq \m-\fr{1}{2}\m_k^{''}(k_*)(k-k_*)^2,
\ee
where $\m_k^{''}(k_*)\simeq 2\m/(k_*)^2$, $\m$ is the average index  depending on the
coupling constant $g$ and it typically varies in-between $0.1$ and $0.2$  \cite{kls2}.  

To calculate the correlation length we use \eq{b} in \eq{c}. As discussed in \cite{kls2}, the integral of  the second term in the square bracket in \eq{c} (the one containing the real part) gives a time dependent oscillating correction which is less than 1 compared to $|\b_k|^2$ integral (see  (89) in \cite{kls2}).
Therefore, \eq{c} can be written as  
\be\label{cc1}
<\chi(t,\bx)\chi(t,0)>\simeq\fr{(1+C)}{2\pi^2 a^3}\int_0^\infty  dk\,\fr{k^2
|\b_k|^2}{\o_k}  \left[\fr{\sin(kr)}{kr}\right], 
\ee
where $C$ is the correction factor mentioned above.  Using \eq{b} and \eq{ma}, the momentum integral in the last equation can now be performed using the steepest-decent approximation, which gives 
\be\label{stp1}
<\chi(t,\bx)\chi(t,0)>\simeq \left[\fr{\sin(k_*r)}{k_*r}\right] <\chi^2>.
\ee
Therefore, the comoving correlation length can be determined as 
\be\label{pre-c}
\xi_c \simeq \fr{1}{k_*}=\fr{1}{\sqrt{gm\Phi_0}}. 
\ee
In this computation, the effects of the modes in other instability bands and the
time evolution of the first band caused both by the redshift of  momenta and the
decrease of the inflaton amplitude $\Phi$, which would lessen the width of the
band, are ignored (note that $k_*$ given below \eq{b} is the physical momentum).
However, as shown in \cite{kls2}, the modes that have been amplified from the
very beginning  become exponentially  larger than the others. Thus, in practice
one can treat $k_*$ as a  comoving momentum scale such that the first
instability band does not change in time and \eq{pre-c} should give a good
estimate for the comoving correlation length of fluctuations.  

It is interesting to compare the size of the  correlation length to the Hubble
radius at the end of the preheating. The broad parametric resonance ends when
$q=g^2\Phi^2/m^2\sim1$ and this can be used to estimate the value of the
amplitude  at that time as $\Phi\sim m/g$. The Hubble parameter can be
 determined from \eq{sol} as
\be\label{h43}
H\simeq \fr{m^2}{gM_p}.
\ee
By comparing the initial and the final values of $q$, the amplitude $\Phi$ can
be seen to decrease
$\sqrt{q_0}$ times, where $q_0$ is the initial value of $q$ at the beginning of
preheating. Then, from \eq{sol1}, one sees that the universe expands $q_0^{1/3}$
times and thus the physical correlation length at the end of broad parametric
resonance is given by
\be\label{pre-cp}
 \xi_c^{phys}\simeq q_0^{1/3}\xi_c=\fr{q_0^{1/3}}{\sqrt{gm\Phi_0}}=q_0^{1/12}
m^{-1}.
 \ee
As a result, \eq{pre-c} and \eq{h43} gives the ratio of the physical
correlation length to the Hubble radius as 
\be
\fr{\xi_c^{phys}}{R_H}\sim q_0^{-5/12}\,\fr{\Phi_0}{M_p}.
\ee
One can have $\Phi_0\sim M_p$, but $q_0\gg1$ in the broad parametric resonance
regime, thus the ratio is in general less than unity.

In this model, the variables $T$, $V$ and $G$ can be determined in terms of
$<\chi^2>$ as follows. Firstly, from \eq{func}, one sees that 
\be
V=\fr{1}{2}M^2<\chi^2>=\fr{1}{2}g^2\Phi^2\sin^2(mt)<\chi^2>.
\ee
Therefore, $V$ oscillates between zero and the maximum value
$V_{max}=g^2\Phi^2<\chi^2>/2$. To determine $G$, one can again use \eq{b} and
\eq{ma} in \eq{func},  and apply the steepest-decent approximation. The steps
are identical to the derivation of \eq{stp1} and one finds 
\be
G\simeq \fr{k_*^2}{a^2}<\chi^2>.
\ee
To find $T$,  one can first ignore the expansion of the universe  to a very good approximation. Using \eq{ab2} in \eq{func} one then obtains 
\bea
T=\fr{1}{4\pi^2 a^3}\int_0^\infty k^2\, \o_k \left[
|\b_k|^2-\textrm{Re}\lf\a_k\b_k^* e^{-2i\int_{t_0}^t \o_k dt'}\rg \right]dk\nn
\eea
where we regularize this expression  by ignoring an additive factor of $1/2$ in
the square brackets. Applying again the steepest-decent approximation, we find  
\be
T\simeq\fr{1-C}{1+C}\o_{k_*}^2<\chi^2>,
\ee
where $C$ is the factor defined in \eq{cc1}. Since from  \eq{om1} $\o_{k_*}$ is an
oscillating function of time, $T$ is also oscillating. However, it is easy to
see that $V+G\simeq T$.  

Finally, it is also possible to determine the phase $\phi_k$, which first
appeared in \eq{dr}. Ignoring the expansion of the universe, $\phi_k$ equals 
the difference between the arguments of $X_k$ and $\dot{X}_k$. Using
$|\a_k|\simeq|\b_k|$, one can see that $\textrm{Arg}(X_k)\simeq \varphi_k/2$ and
$\textrm{Arg}(\dot{X}_k)\simeq \varphi_k/2 +\pi$, where $\varphi_k$ is the phase
difference between $\a_k$ and $\b_k$. Therefore 
\be
\phi_k\simeq \pi.
\ee
From \eq{dr} and \eq{dp}, the value of $\phi_k$ can be seen to favor the energy fluctuations
compared to pressure fluctuations.

\subsection*{Perturbative decay}

Let us now consider the reheating process in perturbation theory. For that  we
assume the following trilinear coupling 
\be\label{int2}
{\cal L}_{int}=-\fr12 \s \phi\chi^2,
\ee
which may arise after spontaneous symmetry breaking.
In this model, the frequency \eq{o} becomes
\be\label{o2}
\o_k^2=\fr{k^2}{a^2}+\sigma \Phi\sin(mt).
\ee
As discussed in \cite{kofrev}, the perturbation theory is applicable when
$\s\Phi/m^2\ll1$, which can in general be satisfied if the amplitude  $\Phi$ is
small.  It is known that the  preheating picture completely changes when both
interactions \eq{int1} and \eq{int2} present in the Lagrangian, see \cite{dfkpp}.
Therefore, the perturbative decay due to \eq{int2} should be considered on its
own as a different model, i.e. it is not to be preceded by the preheating
considered above. Let us remind that the evolution of the background fields is
still given by \eq{sol1}. 

In perturbation theory i.e. for $|\b_k|\ll1$, Bogoliubov coefficients can
iteratively be solved and to first order they can be determined as  
\bea
&&\a_k\simeq1,\nn\\
&&\b_k\simeq \fr12 \int_{t_0}^{t}dt'\,
\fr{\dot{\o}_k(t')}{\o_k(t')}\exp\left(-2i\int^{t'} \o_k(t'')dt''\right).\label{ps} 
\eea
The time integral in \eq{ps} can be evaluated using the stationary phase method \cite{strqc}
(note that in almost all models of inflation $H\ll m$, which is important for the applicability of the stationary phase approximation). There are two  oscillatory terms in \eq{ps}, one is the explicit pure phase exponential and the other is coming from $\dot{\o}_k/\o_k$.  It can be shown that for a given $k$, the main contribution to the integral comes from an interval near  $t_*$ fixed by $\o_k(t_*)=m/2$,
which, in the perturbative regime $\s\Phi/m^2\ll1$, implies  
\be\label{c1}
\fr{k}{a_*}=\fr{m}{2}.
\ee
This can be interpreted as the decay of the inflaton at time $t_*$ to two $\chi$ particles with comoving momentum $k$ (see e.g. \cite{kofrev}).  Thus, only the modes in the following
interval 
\be\label{reg}
\fr{a_0m}{2}<k<\fr{a_1m}{2}
\ee
significantly produced, where $a_0$ and $a_1$ are the scale factors at the
beginning and at the end of the decay, respectively.  We quote from \cite{m1}
the  result of the relatively straightforward calculation:
\be\label{bf}
|\b_k|^2\simeq\fr{\pi\s^2M_p\Phi_0}{2m^{5/2}}\left(\fr{a_0}{2k}\right)^{3/2},
\ee
where $\Phi_0$ is the initial value of the amplitude. 

To determine the correlation length, we use \eq{ps} in \eq{c} and find   
\bea
&&<\chi(t,\bx)\chi(t,0)>=\fr{1}{2\pi^2 a^3}\int_0^\infty  dk\,\fr{k^2}{\o_k}
\times\label{c2}\\
&&\left[|\b_k|^2 +\int_{t_0}^t dt'  \fr{\dot{\o}_k(t')}{\o_k(t')}
\cos(2\int_{t'}^t
\o_k(t'')dt'')\right]\left[\fr{\sin(kr)}{kr}\right].\nn 
\eea
The time integral in the square brackets in \eq{c2} can also be performed using
stationary phase
method. However,  the phase integral now produces an extra cosine term, 
\be
\cos\lf \int_{t_0}^{t_*} \left[2\o_k(t')+m\right] dt'\rg,
\ee
where $t_*$ is the time corresponding \eq{c1}. This term can be seen to
oscillate very rapidly in the decay range \eq{reg} and thus the second term in
the square brackets in \eq{c2} can be neglected since its contribution will be
very small after performing the momentum integral. Using \eq{bf} in \eq{c2} then
gives  
\be
<\chi(t,\bx)\chi(t,0)>\simeq B \int_{a_0m/2}^{a_1 m/2}   dk\,
\fr{\sin(kr)}{k^{3/2}},
\ee
where $B$ is a $k$-independent constant. The indefinite integral  can be
explicitly evaluated in terms of the Fresnel cosine integral $C(x)$  as
\be
2\sqrt{2\pi r }\,\textrm{C}\lf\sqrt{\fr{2rk}{\pi}}\rg-\fr{2\sin(kr)}{\sqrt{k}}.
\ee
One can now see that to a very good accuracy the {\it comoving} correlation
length is independent of the upper limit  of the integral $a_1 m/2$, instead it
is fixed by the lower limit as  
\be
\xi_c\simeq \fr{1}{a_0m}.
\ee
Thus, the physical  correlation length at the end of the decay is  given by
\be\label{per-c}
\xi_c^{phys}\simeq \fr{a_1}{a_0m},
\ee
which is larger than $1/m$ by the expansion factor of the universe during
reheating. 

By noting from \eq{bf} that the spectrum  is given by $|\b_k|\sim k^{-3/4}$, it is easy to understand \eq{per-c}  physically as follows. The dependence of $|\b_k|$ on the comoving momentum $k$  indicates that, as far as the correlation length is concerned, the particle creation effects decrease with increasing momentum. Since  the process  can be thought as the decay of the inflaton particle with mass $m$ into two $\chi$ particles with physical momentum $m/2$, the particles created in the beginning of the process have the smallest comoving momentum $a_0 m/2$ and thus are the most important ones. The corresponding wavelength, which redshifts  in time, gives the correlation length \eq{per-c}. All the particles created in due time have larger momenta and smaller wavelengths, and they produce sub-leading corrections to the correlation length.  

To compare the correlation length to the Hubble radius, let us determine the Hubble parameter at the end of the decay. As discussed, e.g. in \cite{kofrev}, the decay process described by the interaction \eq{int2} is equivalent to decay with a constant decay rate $\Gamma_\phi\sim \s^2/m$, which gives a reheating temperature $T_R\simeq\sqrt{\Gamma_\phi M_p}\sim \sqrt{\s^2M_p/m}$. The Hubble parameter corresponding to this temperature  is given by 
\be\label{hpert}
H\sim\fr{\s^2}{m}.
\ee
Therefore, 
\be
\fr{\xi_c^{phys}}{R_H}\sim \,\fr{a_1\s^2}{a_0m^2}.
\ee
In these models one usually imposes $\s\ll m$ for perturbation theory to be applicable. On the other hand, the expansion factor during reheating depends on  the initial value of the inflaton amplitude, but it is not expected to be a very large number. Thus, one again finds that the correlation length is
smaller than the Hubble radius.  
 
In perturbation theory,  the functions \eq{func} can also be approximately
determined in terms of $<\chi^2>$. By definition the potential energy $V$ is
given by 
\be
V=\fr{1}{2}\s\Phi\sin(mt)<\chi^2>.
\ee
The calculations of the variables $T$ and $G$ proceed as follows. 
Compared to the momentum integral in $<\chi^2>$, the integrand in $T$ and $G$
contains  two more powers of $k$ (note  that in the decay range $\o_k\simeq
k/a$). Evaluating the integrals, the dominant contribution comes from the Ultra-Violet (UV) end
of the limit, which is equal to $m$, and this gives 
\be
T\simeq G\simeq m^2<\chi^2>.
\ee
Since in perturbation theory $\s\Phi/m^2\ll1$,  one finds that  $T\sim G\gg
V_{max}$, where  $V_{max}$ is the maximum value of the potential energy $V$. 

From \eq{ab} and \eq{ab2} the phase $\phi_k$ can also be determined easily when
the expansion of the universe is neglected. Using  $|\a_k|\gg|\b_k|$, one finds that
$\phi_k\simeq 3\pi/2$. As oppose to the preheating, the integrals involving
$\phi_k$ in \eq{dr} and \eq{dp} now tend to decrease $\d\r$ and increase $\d P$.
Finally, by using \eq{bf} in \eq{c} (and taking the $r\to0$ limit)  one can obtain  
\be\label{cp}
<\chi^2>\sim \fr{\s^2\Phi M_p}{m^2}.
\ee
Not surprisingly, one sees from \eq{cp} that in perturbation theory $<\chi^2>$ is smaller by several
orders of magnitude compared to  $M_p^2$,  and it is also much smaller than
$m^2$.  

\subsection*{Thermalization}

After determining the correlation length of quantum fluctuations, let us now
discuss possible effects that they might produce during reheating. We first
consider the thermalization process of the decay products. This is a
difficult process to study and unfortunately there is not much work done in the
literature (see e.g. \cite{t1,t2,t3,t4,t5,t6,t7,t8}). In general, the
distribution of particles produced during the decay of the inflaton can be seen
to be far from thermal equilibrium \cite{t1}. Thus, depending on the interaction
rates  the full thermalization can be delayed resulting a low reheat temperature. 
 Here, we would like to point out that a very quick or instant thermalization of
the decay products may also be prohibited by the existence of quantum fluctuations in the
energy density.

It is known that in thermal equilibrium at temperature $T$ the energy density is
given by 
\be\label{te1}
\r=\cc T^4,
\ee
where $\cc$ is a constant depending on the number of bosonic and fermionic species in  equilibrium. Then, $\overline{U}_V =\r V$, where $\overline{U}_V$ is  the average  energy  inside a physical volume $V$. In the canonical ensemble, the dispersion of the energy about its mean value is given by 
\be\label{te2}
\overline{U_V^2}-\overline{U}_V^2=-\fr{d\overline{U}_V}{d\b},
\ee
which, using \eq{te1}, implies (see e.g. \cite{thermal})
\be\label{th}
\fr{\d U_V}{\overline{U}_V}=\fr{2}{(\cc V)^{1/2}}\fr{1}{T^{3/2}}. 
\ee
In \eq{th}, the dependence of the relative deviation on $V$ is a characteristic feature of thermal fluctuations. 

Thermal equilibrium can only be justified if thermal fluctuations at a given scale is larger than the quantum fluctuations at the same scale. Otherwise, one should wait for some time for the energy to be transferred in space for uniformization.  (see the discussion following \eq{tson} below).   In a volume $V$, there are on the average $V/(\xi_c^{phys})^3$ number of uncorrelated regions as far as quantum fluctuations are concerned. Since in each region the relative deviation of the energy density is of the order of unity, the relative  order of quantum fluctuations in the volume $V$  can be found as
\be
 \fr{\d U_V}{\overline{U}_V}\simeq
\fr{(\xi_c^{phys})^{3/2}}{V^{1/2}}.
\ee
Comparing with \eq{th}, one sees that instant thermal equilibrium after the
decay can only be justified if  
\be\label{te3}
T<\fr{1}{\xi_c^{phys}\cc^{1/3}},
\ee 
i.e. when the thermal correlation length is larger than $\xi_c^{phys}$. Eq. \eq{te3} 
places an upper bound on the reheating temperature in terms of the correlation
length of quantum fluctuations. Note that since thermal fluctuations can occur
locally, the volume $V$ about which the fluctuations are compared should be
taken inside the horizon and thus it is important to have $\xi_c^{phys}$ to be
less than the Hubble radius.  

Since the Hubble parameter should not change abruptly, the final equilibrium
temperature $T_R$ can be determined from the Friedmann equation  
\be
H^2\sim \fr{\r}{M_p^2}\sim \fr{T^4}{M_p^2},
\ee
which gives $T_R\simeq\sqrt{HM_p}$. The full thermalization can be achieved when
$\C_\chi \sim H$, where $\C_\chi$ is the total interaction rate of the decay
products and should not be confused with $\C_\phi$, i.e. the decay rate of
the inflaton.

In the preheating model reviewed above, the Hubble parameter at the end of the  broad
parametric resonance is determined in \eq{h43}. If thermal equilibrium sets in a
very short time following the end of the broad parametric resonance, the reheating
temperature must be fixed as 
 \be
 T_R\sim \sqrt{\fr{m^2}{g}}.
 \ee
Then, \eq{pre-cp} and \eq{te3} give 
\be
g>\cc^{2/3}q_0^{1/6}.
\ee
Since $q_0\gg1$, the coupling constant becomes large $g>1$, which shows that the quantum fields $\phi$ and $\chi$ are strongly interacting. This is a very difficult regime to study and the results about the preheating period reviewed above can no longer be  trusted.  For $g<1$, \eq{te3} cannot be satisfied, which shows that even when the total interaction rate is large to yield quick thermal equilibrium in principle, this cannot be achieved since would be thermal fluctuations are (much) smaller than quantum fluctuations.  

On the other hand,  in the perturbative decay described by the interaction \eq{int2},
the reheating temperature is given by $T_R\sim\sqrt{\s^2M_p/m}$ (here one assumes 
$\C_\phi\sim\C_\chi\sim \s^2/m$).   Then, \eq{per-c} and \eq{te3} imply   
\be\label{tson}
\lf\fr{a_0}{a_1}\rg^2\fr{m^3}{\s^2}>M_p.
\ee
It is not very difficult to satisfy this condition since one usually assumes $\s\ll
m\ll M_p$. 

If \eq{te3} is not satisfied, one should wait for the energy to be redistributed  
for thermal equilibrium. This process is different than the redistribution of
{\it mean} energy or {\it mean} number density by scattering or particle
decays, as analyzed, for instance, in \cite{t4}. Namely, fluctuations should
propagate in space-time to achieve uniformization and this cannot happen
instantaneously because of causality. 

To illustrate what would happen when \eq{te3} is violated, i.e. when quantum fluctuations are larger than thermal fluctuations, assume that $\C_\chi$ is large enough to convert $\chi$ particles into radiation in local thermal equilibrium in a very short time.  Consider now two different regions that have the size of the correlation length and assume that in one region all the energy stored in the inflaton oscillations is converted into $\chi$ particles and in the other region only half of the energy is converted into $\chi$ particles. This is a typical situation since the deviation in the energy density of the created $\chi$ particles is comparable to its mean value.  Let us now try to imagine what would happen as the universe expands twice. In the first region the energy density, being composed of radiation in equilibrium, will be
redshifted  by 16 times, while in the second region half of the energy density, i.e. the radiation, will be redshifted by 16 times and the other half corresponding to inflaton oscillations will be redshifted 8 times. 
Therefore, there appears a genuine density contrast $\d\r$ and one can easily see that because of the different redshifts of the constituent energy densities, $\d\r/\r$ grows with the expansion like $\d\r/\r\sim a$.
 
On the other hand,  the free propagating of radiation is expected to work for the uniformization of energy,   but this process is slowed down by causality. Namely, by looking at the evolution of radial null geodesics, a comoving spherical region of volume $V$ can be seen to expand (in the comoving grid) as $a^{3/2}$. Therefore, $\d\r/\r$ is expected to decrease like $\d\r/\r\sim1/\sqrt{V}\sim a^{-3/4}$ which is smaller than the increase $\d\r/\r\sim a$ noted above. As a result, one sees that the fluctuations in the $\chi$ field induce  changes in the total energy density during the decay process. Later on, when  everything is converted into radiation, the fluctuations tend to be smoothed out, but  depending both on the magnitude and the scale, one should wait for some time for full thermal equilibrium to set in. 

This naive scenario should be sharpened  by considering the evolution of all
fields, especially by taking into account the back-reaction of the created
particles on the inflaton oscillations and geometry. In any case, one
sees that quantum fluctuations can be larger than (would be) thermal
fluctuations in equilibrium and thus they should be taken into account when the thermalization
of the decay products  is studied. 

\subsection*{Back-reaction}

Another interesting process that can potentially be affected by quantum fluctuations is the back-reaction of created particles on the evolution of the background fields. It is important to understand  back-reaction effects to get a complete picture    of the reheating process. Here, we consider the preheating model with interaction \eq{int1} reviewed above. When the created $\chi$ particles  are taken into account, the background scalar field equation in \eq{fe2} should be modified as
\be\label{ff1}
\ddot{\phi}+3H\dot{\phi}+m^2\phi -g^{ij}(\del_i\del_j\phi)+g^2 \chi^2\phi=0.
\ee
One usually invokes Hartree approximation and uses  the expectation value
$<\chi^2>$ for  $\chi^2$ in \eq{ff1}.  Since $<\chi^2>$  does not depend on
spatial coordinates and since initially (i.e. just after the end of inflation)
the $\phi$ field  is homogenous, only the {\it  zero mode} continues to exist 
and one can ignore the spatial derivatives in \eq{ff1}.  In that case, when the mean value of $\chi^2$ grows to satisfy 
\be\label{bchi}
g^2<\chi^2>\sim m^2,
\ee
the back-reaction effects become important. According to \cite{kls2}  this happens when    
\be
\overline{n}_\chi\simeq \fr{m^2\Phi}{g},
\ee
where $\overline{n}_\chi$ is the average total number density of $\chi$ particles.

However, as shown in the previous section,  $\chi^2$ has fluctuations comparable to its vacuum expectation value, therefore as $<\chi^2>$ grows and \eq{bchi} is satisfied, the  "actual value" of $\chi^2$ �depends highly on the position, especially  when it is compared at  scales larger than the correlation length $\xi_c$. To determine  the subsequent evolution of the inflaton zero mode, we imagine the
space to be divided into regions of volume $\xi_c^3$ and analyze  
\eq{ff1} in each  region independently, trying  to predict the full motion by gluing the
results. Note that if only the zero mode survives even when the fluctuations are
taken into account, then this should be a good approximation. 

From \eq{ff1} and by ignoring the expansion of the universe, the frequency of the oscillations in the $i$'th region is given by
\be\label{ox}
\o_{(i)}^2=m^2+g^2\chi_i^2,
\ee
where $\chi_i^2$ denotes the value of $\chi^2$ in that region. Since $\o_i$ is
(nearly) uniform in the $i$'th region, one can approximately write 
\be \label{ap1}
\phi_i\simeq \Phi \sin \lf\o_it\rg,
\ee
where $\phi_i$ denotes the restriction of $\phi$ in that  region. 

Let us now try to see if these local solutions, which are valid in different regions, can smoothly be glued. From \eq{ox}, one sees that when $g^2<\chi^2>\sim m^2$, the frequencies start to differ  from one region to another as
\be
\fr{\d\o}{\o}\sim \fr{\d\chi^2}{\chi^2}\sim 1.
\ee
Therefore, in time $t\sim 1/\o\sim1/m$, nearly corresponding to a single
average oscillation, the oscillations  of the inflaton field in different regions 
become completely  out of phase. 

Assume now that $\phi=\sum_i \phi_i$, where
each $\phi_i$ has its support in the $i$'th region.  In that case, one can estimate 
the spatial derivatives in \eq{ff1} as
\be
\del_i \phi\sim\fr{\Phi}{\xi_c}, \hs{5}
g^{ij}(\del_i\del_j\phi)\sim \fr{\Phi}{(\xi_c^{phys})^2}.
\ee
By comparing with the mass term, the spatial derivatives can only be ignored if
\be\label{cb}
m^2|\phi| \gg\fr{\Phi}{(\xi_c^{phys})^2}.
\ee
It is not possible to satisfy \eq{cb}, when the  inflaton passes through its
minimum $\phi=0$. When it reaches its maximum $\phi=\Phi$, spatial derivatives
can be neglected if 
\be\label{85}
m\gg\fr{1}{\xi_c^{phys}}.
\ee
Using \eq{pre-cp} as an estimate for $\xi_c^{phys} $,  \eq{85} requires
$q_0^{1/12}\gg1$, which can only be satisfied if $q_0\gg 10^6$, or so. As a
result, one sees that when back-reaction effects become important the coherence
of the inflaton oscillations is lost and the spatial derivatives  must not be
neglected. 

If preheating continues after back-reaction, our findings indicate a major
change in the whole process. First of all, with the inclusion of spatial
derivatives it becomes much more difficult, if not impossible,
 to determine the the subsequent evolution of the inflaton field. Moreover, one
should also revise the particle creation process  since the evolution of the 
background drastically changes. Presumably, it will no longer be possible to
talk about resonance bands in momentum space, but instead it would be more
appropriate to analyze the process region by region. 

\subsection*{ Formation of primordial black holes}

Finally, in this subsection we point out  a new mechanism for the formation of
primordial black holes, which can be efficient especially in models of
preheating. As we will see, this process is different than the one considered,
e.g. in \cite{bhc}, where a sufficiently large density contrast, whose amplitude  is
greater than a critical value as it enters the horizon, dynamically evolves in
time to form a black hole. 

We show in the previous section that the energy density $\r$ has fluctuations
about its mean value. This is not surprising,  since in quantum theory \eq{ex} 
implies that $\r$ becomes a random variable as it is a function of a  Gaussian
random variable $\chi$. In general one can try to determine (at least an
approximate) probability distribution for $\r$, which must be close to a 
chi-squared distribution, but this is in general a difficult task since it
contains two "non-commuting" random variables $\chi$ and $\dot{\chi}$. Instead
of considering the full energy density, in the following we focus on the potential energy part
\be
\r_P=\fr{1}{2}M^2\chi^2,
\ee
which has a simple probability distribution determined in terms of the Gaussian
distribution of $\chi$. It is clear that $\r>\r_P$ and if  $\r_P$   exceeds the
critical value for a black hole to form then so does $\r$. 

Consider now the potential energy $E_P$ stored in a region of size
$\xi_c^{phys}$, which is given by
\be\label{ebh}
E_P=\r_p (\xi_c^{phys})^3\simeq M^2\chi^2 (\xi_c^{phys})^3.
\ee
If the Schwarzschild radius  corresponding to this energy becomes greater than
$\xi_c^{phys}$, then a black hole must form in that region (as discussed in the
previous subsections, $\xi_c^{phys}$ is in general smaller than the Hubble
radius and thus it is legitimate to ignore  the expansion of the universe).
Recalling that the Schwarzschild radius for  the energy $E$  is given by
$R_S\simeq E/M_p^2$, the condition  $R_S>\xi_c^{phys}$ for the energy \eq{ebh}
becomes
\be\label{bhf}
\chi^2>\chi_0^2\equiv \fr{M_p^2}{M^2 (\xi_c^{phys})^2}.
\ee
Since $\chi$ is a Gaussian random variable with zero mean,  the normalized
probability distribution can be written as 
\be\label{dist}
{\cal P}(\chi)=\fr{1}{\sqrt{2\pi <\chi^2>}}\exp\lf-\fr{\chi^2}{2<\chi^2>}\rg.
\ee 
Therefore, the probability of getting $\chi^2>\chi_0^2$, which is also the probability for a black hole to form,  is given by
\be\label{p}
{\cal P}=2\int_{\chi_0}^{\infty} P(\chi)\,d\chi.
\ee
Note that, we underestimate this probability, since only
the contribution of the  potential energy is considered. Nevertheless,
\eq{p} should give a good estimate since one would not expect a large
hierarchy between the potential and total energies. 

On the other hand, because of the time dependence of  the variance $<\chi^2>$ (and in
some cases $\chi_0$), the probability distribution \eq{dist} and the probability
\eq{p} changes in time. In that case, the black hole formation process can be visualized
in time as follows.  Assume at a fixed time $t$, some ${\cal P}(t)$ fraction of regions
collapse to form black holes. Then one should wait for some time for the field
$\chi$ to evolve  according to the new probability distribution and then regions
containing large $\chi^2$ fluctuations collapse again. Therefore, the process is
actually cumulative. Moreover, the energy density in a collapsing region is
larger than the average energy density in the universe, thus the actual fraction
of energy that goes into primordial black holes must be larger than ${\cal P}$.  All
these arguments support the use of \eq{p} as a conservative estimate for black hole formation probability.

If $\b$ denotes the observational upper limit of the fraction of energy that can
go into the primordial black holes, then one should impose 
\be\label{conbh}
\b>{\cal P}.
\ee
It is known that  $\b<10^{-20}$ (see, e.g. \cite{bhc}). The magnitude of ${\cal P}$ depends very sensitively on the ratio  $<\chi^2>/\chi_0^2$ and numerically one can find that to satisfy \eq{conbh} with a (small) margin the following condition must be obeyed: 
\be
\fr{\chi_0^2}{<\chi^2>}>22.
\ee
Using the definition of $\chi_0^2$ from \eq{bhf}, this implies
\be\label{son}
\lf<\chi^2>\rg<\fr{M_p^2}{22M^2 (\xi_c^{phys})^2}.
\ee
Therefore, in a model giving a large vacuum expectation value for $\chi^2$ and a
small correlation length $\xi_c^{phys}$, formation of primordial black holes can
be a problem. It is clear that this can be a  dangerous issue especially for preheating. 

Let us check, for example, if  the model \eq{int1} reviewed above passes the
condition \eq{son}. Assume that the broad parametric resonance ends just when
the back reaction effects become important, i.e. $q=1$ when $g^2<\chi^2>\sim
m^2$. As shown in \cite{kls2}, for the realistic choice of the inflaton mass
$m=10^{-6}M_p$, this happens for the coupling $g\simeq 3\times10^{-3}$ and
$q_0\simeq 10^6$. Let us try to evaluate terms in \eq{son} in terms of
$g$.  First of all, from \eq{om1}, the mass parameter $M$ can be seen to be
oscillating with an amplitude $g\Phi$. So, we take $M=g\Phi$ in \eq{son}, i.e.
we consider the  black hole formation when the inflaton reaches its maximum
value. As noted above, the vacuum expectation value can be estimated  as
$<\chi^2>\simeq m^2/g^2$ and the amplitude  $\Phi$ can be found from $q\sim 1$
as $\Phi\simeq m/g$. Combining all these information, one sees that \eq{son}
implies $g>1.5 \times 10^{-5}$, which is satisfied with a good margin since the actual
value of the coupling is taken as $g\simeq  3\times10^{-3}$. Thus one concludes
that formation of black holes is not an issue in this model.  

\section{Conclusions}\label{ch4}

In this paper, we consider the well known particle creation effects in a time-dependent, 
homogenous and isotropic, classical background and point out a feature that has
not been elaborated in detail previously. Namely, we examine the fluctuations of
important physical quantities characterizing the particle creation
process about their vacuum expectation values. We specifically consider  the
energy density, the pressure and the other components of the energy-momentum
tensor, and find that all these quantities have in general large fluctuations
comparable to average energy density. We also note that the deviation
corresponding to the field square $\chi^2$ also equals $\sqrt{2}$ times
the mean value of $\chi^2$. It is possible to make sense of  all these fluctuations
by using adiabatic or WKB regularization, therefore they have direct and
unambiguous physical meaning. 

The spatial scale of these fluctuations is given by the correlation length of the quantum field $\chi$ excited by the classical background. This is an important observation, which allows one to think of these fluctuations being defined  in  regions that have the size of the correlation length, instead of imagining them as independent random variables defined at different points. If one still maintains the view that the field variables are defined point-wise, then one should always keep in mind that the variables inside the same region are correlated with each other and behave in the same way. 

In the second part of this paper, we focus on the reheating process in single
scalar field driven inflationary models and investigate possible implications of
our findings. We consider two well known models, which are typical examples of
the decay in broad parametric resonance regime and in perturbation theory, and
determine the correlation length of quantum fluctuations. In both models, the
physical correlation length becomes smaller than the Hubble radius at the end of
the decay, where the ratio tends to be smaller in perturbation theory. 

We investigate three possible effects of these fluctuations during reheating period. The first one is related to the final thermalization  process of the decay products. One usually assumes that  the decay products  reach thermal equilibrium in a very short time, which is possible if the interaction rates  are  comparable to the expansion rate at the end of the decay.  Although,  this assumption is difficult to satisfy in realistic scenarios  (see, e.g. \cite{t4}), it is important to estimate the maximum reheating temperature in a given model. We show that even when the interaction rates are presumed to be large enough and  thermal equilibrium is expected to be set  in a very short time, the existence of quantum fluctuations may delay this process. Namely, we observe that if on a given (subhorizon) scale quantum fluctuations are larger than would be thermal fluctuations in equilibrium then one should wait for energy to be redistributed in space to obtain the real equilibrium. Especially for larger temperatures, the thermal correlation length can be many orders of magnitude smaller than the quantum correlation length, which requires more time for energy to spread.  On the other hand, the whole process can actually be much more difficult to study since back-reaction effects must be considered to obtain the complete picture.

As a second event which might be affected by quantum fluctuations,  we consider the back-reaction of the created particles on the inflaton oscillations in the model of preheating. Back-reaction becomes important when the mean value of $\chi^2$  sufficiently grows to modify  the frequency of the
inflaton oscillations given by the inflaton mass. When this happens, because of the fluctuations in $\chi^2$ the  oscillation frequency starts to appreciably vary on scales larger than the correlation length.  We show that in  a very short time the coherence of the oscillations is lost and spatial derivatives of the inflaton can no longer be neglected  in determining its dynamical evolution.  From that moment on,  the whole  back-reaction process changes and the problem becomes much more difficult to study since it involves non-homogenous fields in space. 

Finally, we notice that a large fluctuation in the energy density, which occurs in a region that has the size of the correlation length, causes a black hole to form if the corresponding Schwarzschild radius becomes greater than the correlation length. To our knowledge, this process is considered for the first
time in the context of reheating as a plausible mechanism for the formation of  primordial black holes. In previous studies  (see e.g. \cite{bhc}),  the collapse of a sufficiently large, horizon size density contrast, which is known to produce a black hole as a result of its dynamical evolution under the influence of classical gravity, is considered as the main mechanism. We estimate the fraction of energy that can go into primordial black holes because of large energy fluctuations and show that observational constraints impose some new restrictions for the models.  

It is clear from our findings that, as far as the  reheating after inflation  is concerned, the correlation length is one of the most important parameters  characterizing the quantum fluctuations in the particle creation process. While sometimes  a larger correlation length amplifies the impacts of quantum fluctuations as in  the final thermalization process, in some instances a smaller correlation length can increase the effects; formation of primordial black holes being an example as discussed below \eq{son}. 

It would be interesting to extend the present work in different directions. For example, it is of interest  to study the back-reaction effects in more detail in the preheating model. Particularly, new physics may arise if the broad parametric resonance continuous to exists when the back-reaction effects become important. It would also be interesting to extend the fluctuation analysis to include perturbations of 
the other fields, i.e.  the metric and the inflaton. The fluctuations in pressure  and other components of the energy-momentum tensor are expected to play a role in this study. Finally, it is clear that the existence of these fluctuations should be considered when  symmetry breaking effects or formation of topological defects are studied.  For example,  in \cite{cs}  formation of cosmic strings in a preheating model is studied by solving classical field equations numerically.  Although in a linear theory such as considered in \cite{cs},  the classical field equations determine the evolution of the mean values,  it would not be surprising to see the  destruction  of cosmic strings  longer than the correlation length because of the existence of large, order one  fluctuations. It would be interesting to check this expectation by an explicit computation.

\end{document}